\documentclass[prb,preprint,showpacs,aps]{revtex4}

\usepackage{graphicx}

\begin{document}

\title
{Electronic structure and the minimum conductance of a graphene layer on  SiO$_2$ 
from  density-functional methods.}

\author
{ M.W.C. Dharma-wardana
}
\email[Email address:\ ]{chandre.dharma-wardana@nrc.ca}
\affiliation{
National Research Council of Canada, Ottawa, Canada. K1A 0R6\\
}

\date{03 March 07}
%
%
\begin{abstract}
The effect of the SiO$_2$ substrate on a graphene film
is investigated using realistic but computationally convenient
energy-optimized  models of  the substrate supporting a layer
of graphene. The electronic bands 
are calculated using density-functional methods 
for several model substrates. This provides an estimate of the
substrate-charge effects on the behaviour of the bands near $E_F$,
as well as a variation of the equilibrium distance of the graphene sheet. 
A  model of a wavy graphene layer is 
examined as a possible candidate for understanding the
nature of the minimally conducting states in graphene.
\end{abstract}
%
\maketitle
\section{Introduction.}
Ideal graphene is a two-dimensional (2D) sheet of carbon
atoms arranged on a honeycomb lattice.
Given the already rich physics of
graphene, a tremendous effort is being focused on its
basic physics as well as its technological applications\cite{geim,novos,zhang}. 
These include applications based on carbon
 nanotubes\cite{dress,cdw}, or structures based on
 graphene itself\cite{takis,phavou}.
  Unlike carbon nanotubes (CNTs) which may be semi-conducting 
or metallic, pure graphene is a 2D zero-gap material having electron 
and hole mobilities similar to those of CNTs\cite{phavou}.
The honey-comb structure with two C atoms per unit cell 
has two degenerate Fermi points at the {\bf K} and {\bf K$^\prime$}
points of the hexagonal Brillouin zone. The conduction and
valance bands touch at the Fermi points,
with linear energy dispersion, implying a zero-mass
Dirac-Weyl (DW) spectrum for very low excitation energies. This leads to
new physics which is strikingly different to that of typical 2-D
electron systems. Thus an unusual quantum Hall effect, and also
 a $\pi$-phase shift
in the de Hass-Shubnikov oscillations have confirmed the
 Dirac-Weyl spectrum.
 The existence of a  ``minimum conductivity (MC)''
 has also been claimed, although
it is not clear if this is a ``universal'' MC, or a sample dependent 
quantity\cite{novos,zhang}. 
%
The density of states in the DW spectrum falls
to zero at the Fermi energy, and hence the conductance should drop to zero
as the gate voltage $V_g\to 0$. Instead, the conductance reaches a
saturation value $\sigma_m$ for small gate voltages $V_g< v_m$,
 with $v_m\sim$ 0.5 eV. Although this may appear to contradict the
behaviour expected from the DW-like effective Hamiltonian defined
near the {\bf K} points of the idealized 2-D lattice, the neglected
features of the real graphene sheet need to be considered when we
consider the $V_g\to 0$ limit. The graphene sheets are not perfect
2-D systems, but are supported on a SiO$_2$ surface. These
are locally crystalline or amorphous structures containing charged
atoms, and their theoretical structural description
 is quite complex\cite{pierre}. 
The charged sites on these substrates
may have an effect\cite{dvm,gali}
on the DW spectrum that becomes crucial as $V_g\to 0$. Many authors have
examined the effect of models of charged centers using Thomas-Fermi screenings models,
Boltzaman or Kubo-Greenwood conductivity theories\cite{shonAndo,Katsnelson,nomumac}
as well as other methods.
While these methods exploit convenient, simple theoretical
methods, they lack an attempt to confront the microscopic details
of the graphene layer and its interactions with the SiO$_2$ substrate. 
While such a detailed picture may not be needed for many purposes, 
it is clearly important to develop atomistic models of the
 graphene-substrate which go beyond linearly screened
structureless scattering models. Thus one aim of this study is to
examine simple, yet atomistically realistic models of graphene on SiO$_2$
substrates which can be handled by first-principles calculations.
Even if we assume that the graphene sheet does not 
interact strongly with the substrate due to adsorbed
atmospheric N$_2$ layers in
between the graphene and the SiO$_2$, or if the graphene film is considered
suspended in space\cite{mgknbr},
%
the assumption of an ideally flat graphene sheet is clearly untenable.
Ideal 2D layers (which are not part of a 3D structure) are expected to be
unstable for a variety of theoretical reasons\cite{peierls, mermin}.
Even a 3-D solid at nonzero temperatures acquires vacancies and 
lattice defects to gain entropy and minimize its free energy. Hence
some authors\cite{pgcn,balatsky}
 have considered that graphene is a disordered system with a distribution
of vacancies. One may consider that the  effect of the
vacancies may be used to blunt the behaviour near
the ideal {\bf K} points. However, our calculations\cite{grp2,sscdw}
for vacancies in graphene show that the energy costs of breaking the
$\sigma$-bonding network are too great to allow any significant vacancy
formation. In fact, the demonstration by Meyer et al\cite{mgknbr}
 that
 graphene sheets are intrinsically wavy provides an important
key to the properties of free-standing or SiO$_2$-supported 
graphene. Here we note that pyrolitically prepared graphene on SiC is
multi-layered and  protects the active
graphene layer from the effects of the substrate\cite{vfhln}.

In section II we present band calculations for a graphene
sheet positioned on model SiO$_2$ surfaces. One model of the substrate
surface  is full of dangling bonds, while the other is saturated;
but both surfaces have charged centers. These calculations show that
$E_F$ is modified by the variations in substrate structure.
These calculations are reviewed in the
context of a wavy graphene sheet as it brings in the statistical
fluctuations of the position of the sheet in the z-direction. 
These provide a simple understanding of the minimum conductance.
%

%
\section{Density-functional calculations of graphene/substrate systems}
Simple tight-binding methods (TBM) or the even  more restricted
Dirac-Weyl model could be successfully
exploited within a limited energy window for pure graphene.
The two sublattices of the bi-partite graphene lattice become
 inequivalent if
vacancies are introduced. Vacancies rupture the $\sigma$-bonding
network and require energies such that elementary models become
unreliable. In a previous paper \cite{grp2} we presented calculations
for vacancies in graphene (at concentrations of $\sim$ 3\% and above) and
showed that the DW model does not
even hold for such systems. Thence we concluded that graphene films
used in quantum Hall studies must be very high quality flakes
free of vacancies. The experimental results of Mayer et al.\cite{mgknbr}
are in agreement with this point of view.
Hence an understanding of other possible effects
on the electronic structure is needed. One possibility 
is the effect of the electrostatic field of charged sites present in the
SiO$_2$  surface. The quasi-cristobalite like
SiO$_2$ is a very versatile structure where bond angles, bond lengths etc.,
can take a variety of values to easily fit in with the chemical environment.
In earlier times, such systems were presented using ``random tetrahedral networks''
and other phenomenological models, or simply ignored in electron-gas like
field-theory models. In Ref.~\onlinecite{pierre} we developed a first-principles
Car-Parinello optimized model for such systems, and having the
capacity to explain core-level and
photo-emission data as well as many other properties. However, adaptation of such
a model to graphene is quite demanding as the simulation cells which
contain both graphene and the SiO$_2$ substrate need to be commensurate with
the lattice vectors of both systems and hence contain  many atoms.
 Hence, given the complexiety of the
SiO$_2$ system itself, we aim to develop simpler models which exploit the
bonding versatility of SiO$_2$ and retain the essential physics. Using the
very stable SiO$_2$ unit we construct a stable SiO ring structure which
provides a simple substrate with charged oxygen and Si centers. These systems
need {\it no} passivating H atoms. In fact, by attaching H atoms we can simulate
the {\it presence} of dangling bonds which form dispersive bands near
the Fermi energy. These even have linear dispersive regions and suggest
interesting possibilities (which would not be discussed here).

\begin{figure}
\includegraphics*[width=8.0cm,height=10.5cm]{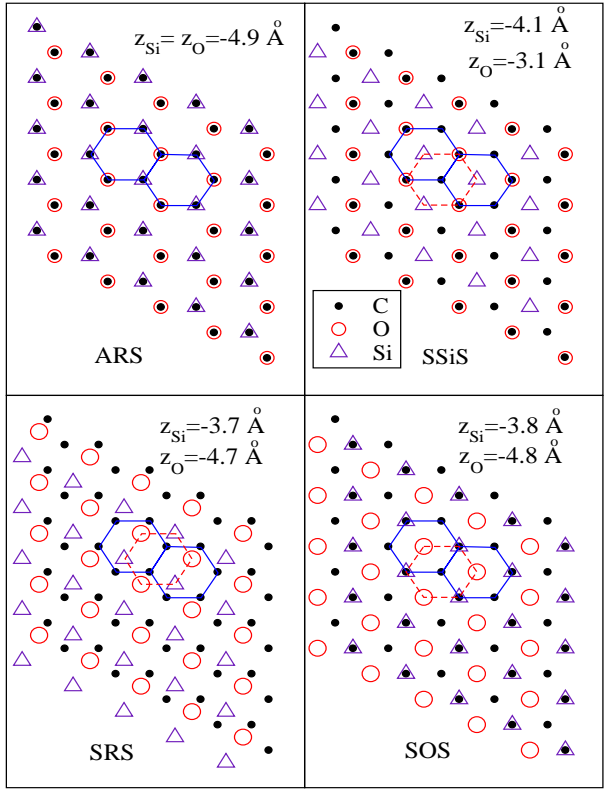}
\caption
{(Color online)The on-top view of the C, Si, and O atoms
 for the structures
ARS, OSS, SiSS and SRS, as described in the text. The O and Si
atoms are below (-ve $z$) the graphene ($x-y$) plane.
 The $z$ distances 
are indicated in each panel.
In H-containing structures, the H-atoms are directly
underneath the Si-atoms.}  
\label{topview}
\end{figure}

\begin{figure}
\includegraphics*[width=8.0cm,height=5.0cm]{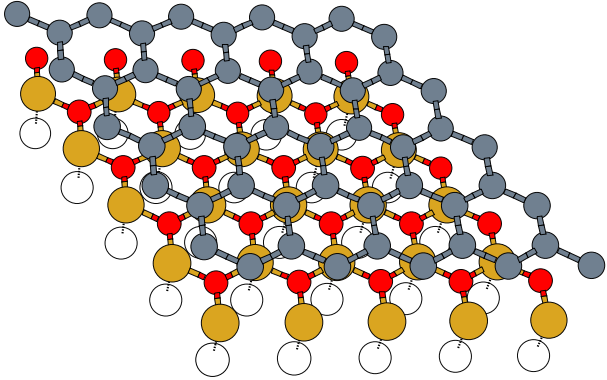}
\caption
{(Color online) Three dimensional view of the structure ARS
with H-atoms attched to Si, and  where the
SiO hexagone is aligned underneath the carbon hexagons.}
\label{structB}
\end{figure}

Thus we consider a {\it stable} 2D Si-O ring structure which satisfies
the lattice vectors of the honeycomb structure 
(see Figs.~\ref{topview},\ref{structB}). In bulk SiO$_2$ the Si-O bond
length is $\sim$ 1.6-1.7 \AA$\,$ and the bond angles are optimal at 106$^o$,
although smaller and bigger angles from 96$^o$-126$^o$
are seen near Si/SiO$_2$ surfaces\cite{pierre}.
Thus there is considerable freedom with respect to bond angles and lengths.
In the structures developed here, bond lengths ranging from $\sim$ 1.43-1.75\AA
are found (on total-energy minimization) as stable structures. The longer bond
length is accommodated by SiO$_2$ layers which are puckered.
 
 The structure with valencies $Z$= 4, and 2 (and two lone pairs) for Si and O is
similar to a charge-transferred N-N system where $Z$=3. The N-N system has no
charge centers, where as the Si-O system has positive charges on Si and negative
charges on O.  It turns out that there
are several possible structures which are in their energy minima. They are 
useful as models for the interaction of the graphene sheet with a SiO$_2$
substrate. We have studied the following cases, depicted
in Fig.~\ref{topview}: (i) The substrate is
represented by a single stable sheet of SiO rings where the Si and O atoms sit
(in the $z$-direction) underneath the C atoms, with the carbon hexagons
{\it aligned} with the Si-O rings; this will be called the
 `aligned ring structure'
 (ARS). (ii)The Si atoms are under the C
atoms and close to the graphene plane, while the O are below the center of the
 carbon hexagons and further away; this structure will be called the
`staggered oxygen structure' (SOS). (iii)The O atoms are under the C
atoms and close to the graphene plane, while the Si are below the center of the
 carbon hexagons and further away from the graphene plane; this is a 
`staggered Si structure', (SSiS). (iv)The Si-O rings are fully staggered from the
carbon hexagons; i.e., a staggered-ring structure (SRS). (v) The ARS with H atoms
attached to the Si-atoms and optimized to give this set of substrate models.
(vi)H atoms attached to the Si in  the SSiS structure and
optimized to give a stable structure.
Unlike the SOS and SSiS structures, this
system treats the bi-partite sublattices equivalently.
The SRS structure is metastable, and 
relaxes to SOS or SSiS if annealed using molecular dynamics,
thus splitting the valley degeneracy. 

Two positionings  (z-locations) of the graphene sheet are studied for the
ARS structure. These structures are typical of a SiO surface in that
they present charged ionic centers (O and  Si), as well as ``dangling''
bonds which are found in  the structures with  H atoms. That is, in these
structures H atoms play a role quite different to the passivating role
played by them in many model structures.
A statistical average over many such possible structures 
is expected in a laboratory sample. The fully microscopic
calculation used here can be used within a numerical averaging
over a large statistical sample of structure realizations. 
However, since we are studying preliminary models, such a
step would be premature. Hence we examine
a few structures which provide some insight into the
graphene-substrate interaction. 

We have used the Vienna {\it ab initio} simulation package (VASP)\cite{vasp}
which implements a density functional periodic plane-wave basis
solution where the ionic coordinates are also equilibriated
to negligibly small Hellman-Feynman forces. The projected augmented wave (PAW) 
pseudopotentials\cite{vasp} have been used for the carbon potential.
This C pseudopotential has already been used with confidence in several 
 graphene-type calculations 
(e.g., Refs.~\onlinecite{grp2, nieminen}). The usual precautions
with respect to $k$-point convergence, size of the simulation
cell, the extent of the plane-wave basis etc.,
 have been taken\cite{grp2} to ensure convergence.  
\subsection{Bandstructure in the presence of the substrate.}
\begin{figure}
\includegraphics*[width=9.0cm,height=12.4cm]{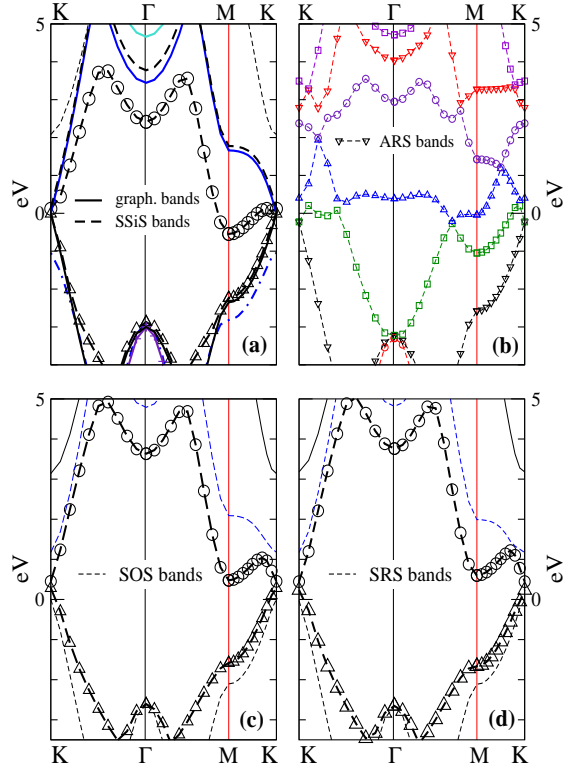}
\caption
{(Color online)Kohn-Sham Energy bands along the {\bf K $\Gamma$ M K}
 directions
of the graphene Brillouin zone, inclusive of substrate effects.
(a) The thick lines are from pure graphene. Dashed lines with
or without data points are for graphne on a SiO substrate with
the O atoms under a C atom. The Si atoms are staggered to the
 center of the C-hexagons (SSiS).
(b)the flat Si-O rings are aligned directly under the C-hexagons (ARS) and
the corresponding bands are shown.
 In panel (C) The Si atoms are under a C atom, and the O atoms are staggered
to the center (SOS), and the resulting bands are shown. The bands in panel (d)
are very similar to those in (c), and is the case where the whole
SiO ring is staggered with respect to the C-hexagon (SRS). $E_F$ is set
to zero in all panels.
}
\label{siobands}
\end{figure}

In Fig.~\ref{siobands}, we show
the band-structure of several graphene-SiO systems
(projected on the graphene Brillouin zone)
along the K$\to\Gamma\to$M$\to$K symmetry directions.
The bands for the structures identified in Fig.~\ref{topview}
as SSiS, ARS, SOS and SRS, and {\it without} any H-atom
attachments, are shown in the figure. The pure graphene bands are
also shown (as continuous lines) in panel (a).
It is clear that the fully aligned flat SiO structure is
not an acceptable model for the graphene substrate. The
other three models largely preserve the linear dispersion 
in the  K$\Gamma$ direction. However, the behaviour near
the {\bf K} point, towards the {\bf M} direction is affected by the 
intrusion of a flat band. The bands shown in panels (a), (c)
and (d) also have a small ($\sim$ 0.15 eV) energy gap which
cannot be seen on the scale of the plots.
However, as we discuss below, the Fermi energy itself moves
by about $\pm$ 0.6 eV from structure to structure and hence
this band-gap splitting is not significant in a real substrate
where many local orientations (of the SSiS, SOS, SRS types) could occur.

\begin{figure}
\includegraphics*[width=6.0cm,height=8.0cm]{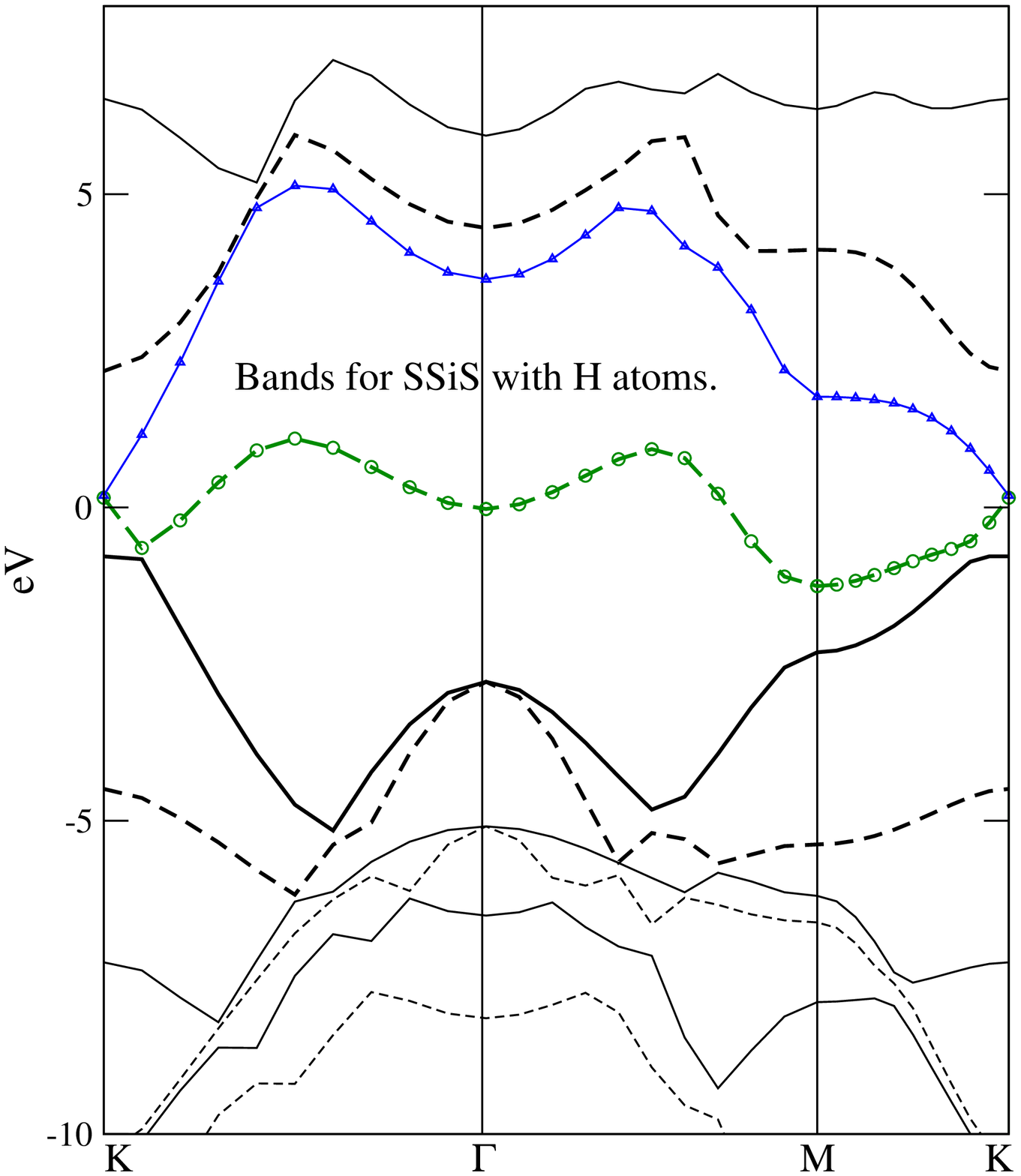}
\caption
{(Color online)Kohn-Sham Energy bands along the {\bf K $\Gamma$ M K}
 directions
of the graphene Brillouin zone, with graphene on a
SSiS substrate where the Si atoms carry H atoms.
The dashed line with circles as data poins is a new
band arising from the H atoms. $E_F$ is set to zero.
}
\label{hbands}
\end{figure}

 In Fig.~\ref{hbands} we show the bands of the SSiS system 
with an H atom attached to the Si atoms. This introduces
an additional band criss-crossing the Fermi energy ($E_F$ is set
to zero in the figure), and marked as a dashed line with
circles.  Hence the H-atom system is not an acceptable model
for graphene on a SiO substrate.
Even the more satisfactory models, i.e.,  SSiS, SOS and SRS shown in
Fig.~\ref{siobands} show that the graphene layer interacts 
with the substrate in a significant manner. Thus, the interaction
between the graphene and the substrate is clearly moderated,
perhaps by adsorbed atmospheric gases, and gives rise to the
almost pure-graphene like behaviour experimentally observed
for graphene films positioned on silica substrates. 

If we consider an extended  substrate including
several SiO layers, as was studied by us in a different
context (as in Ref.~\onlinecite{pierre}),  the 
charge structure of the oxygen atoms may be different in
different regions. 
The different staggered structures present
 local configurations  which would occur in different parts of a single
substrate/graphene setup. The change in the $E_F$ in going
from one part of the substrate where one structure (e.g.,
 the staggered oxygen 
structure) prevails, to  another where another structure
(e.g., the staggered-Si structure) prevails locally, would
 be a measure of the local fluctuations on the graphene
bands caused by the substrate. The calculations for the
Si-O systems shown here, as well as other similar systems
not discussed here, give us an $E_F$ variation $|\Delta E_F|$.
Moving the graphene film from $z\sim$3\AA $\,$ 
to $z\sim 5 $\AA $\,$ changes $E_F$ insignificantly,
although there are changes in the bandstructure. Changing
the alignment of the C-hexagons from the Si-O hexagons
produces a change of $\pm$ 0.6 eV in $E_F$, while the attachment
and detachment of H atoms also produce effects of the same order. 
These calculations suggest that, when the Si-O substrate
modifies the graphene bands without disrupting the DW behaviour,
it could still lead to a spatially varying Fermi energy
 $E_F(\vec{r})$,
with a variation of about
$|\Delta E_F|\sim$ 0.6 eV. We also note that the equilibrium
separation between a large graphene sheet and the substrate
would vary with the {\it local} substrate structure.
Thus, in effect, the graphene sheet would not be flat but
determined by the structural features of the substrate and
the resultant interactions. It should also be noted that a
sheet of graphene simply ``put'' on a substrate does not necessarily
assume the lowest energy conformation, unless an annealing process
is performed. Instead, the film would have a number of ``touch points''
where it would approach the substrate to some near-optimal distance (3-5 \AA$\,$).
These ``touch-points' would hold the film on the substrate, while the
rest of the film could be at other distances. The optimal distance of
3-5 \AA$\,$ is that obtained from the total energy at zero temperature,
and not from the free energy, at the ambient temperature.
Hence the {\it average} distances of up to 8\AA$\,$, or other values,
quoted in experimental studies could easily occur in different samples.
nd for improved models of the SiO$_2$ system.

Our calculations, unlike linear response
or Thomas Fermi models, take account of the structural
features of graphene, the essential features of the Si-O substrate,
as well as the non-linear response,  bound-state structure
and bond-length modifications,
via the self-consistent Kohn-Sham calculations used
here. However, just as the Si/SiO$_2$ interface structure
of field-effect transistors required many decades of study,
the graphene/SiO$_2$ interface would also require more microscopic
calculations. 
\section{Discussion}
A single-sheet of graphene gently placed on a SiO$_2$ substrate
or suspended in space is driven to deviations from perfect flatness
as a consequence of well known 
thermodynamic constraints\cite{mgknbr,peierls, mermin}.
Hence let us consider the wavy-graphene-sheet (WGS) model of
Mayers et al.\cite{mgknbr} and examine the behaviour of such a system in
the low gate-voltage limit. First we note that the DW effective hamiltonian
is obtained by limiting the behaviour to the neighbourhood of $k_F$ which
is at the {\bf K} points. Thus the low-energy excitations are actually
associated with a high-$k$ Fourier component of the system. That is, the 
wave-like elastic energy perturbations of the $\sigma$-bonding
 skeleton (which are due to length scales
of the order 1:10 in $l_z:L_{xy}$, where the in-plane length scale
$L_{xy}$ is  $\sim$ 10 nm), would have little effect on the 2.7 eV
energy scales associated with the large-$k$ regime (i.e., near the {\bf K} 
point) effective Hamiltonian of the DW
spectrum. That is, various elastic-deformation  models based on
extensions of the tight-binding model (fitted to the electronic excitations
near the {\bf K} point) will totally fail to capture the effect of
the undulations of the graphene sheet. However, this point of view is not shared in
Ref.~[\onlinecite{netokim}] where a Slater-Koster tight-binding model, using
the hopping matrix  elements and their spatial derivatives is used. However,
our experience is that such models usually fail to reproduce even the phonons in
the structure. That is, the tight-binding parameters which give good electronic
bands, when used in a phonon calculation do not give good phonons,
 and {\it vice versa}.

In our previous calculations we showed that
the behaviour near the {\bf K} point may survive significant
charge center effects, with the {\it proviso} that the absolute value of
$E_F=\epsilon$({\bf K}) bobs up and down. We found that the $E_F$ for the
various SiO substrates (without attached H) differed by $\sim 0.6$ eV.
Those calculations
assumed the existence of lattice periodicity in each model, with the 
graphene sheet located at $Z_g$=3-5 \AA$\,$, depending on the
substrate configuration. The configuration of the substrate presented to the graphene
sheet (i.e., whether it is SSiS, SOS, or SRS) will differ more or less
randomly, at different ``touch-points''. Thus the {\bf K}-point energy
at each touch-point will differ, with a variation of about $\sim 0.6$ eV.
A statistical averaging over an ensemble of such models can be
carried out, inclusive of charge-charge correlations, using 
methods well known in astrophysics\cite{chandrasekar}, plasma
 physics\cite{microf}, and in recent discussions of random charge 
centers acting on graphene\cite{gali}.
However, we will proceed without such details as the
SiO$_2$ models used here are rather preliminary.
We may note 
that there are several effects to consider:\\
(i) Break-down of the lattice periodicity $a$ and the zone-edge
 periodicity $2\pi/a$.\\
(ii)The modification of the energy at the zone edge and specifically at {\bf K}.\\
Item (i) may be visualized as a replacement of the periodicity $a$ by an
ensemble of periodicities $<L_a>$, and a set of {\it folded bands} with 
zone edges at $2\pi/<L_a>$. The distribution $L_a$ would be such as to minimize
the free energy of the graphene film in the field of the substrate,
or floating free in space. In practice the film is ``put'' on the
substrate and its conformation may be some metastable one and not necessarily
the lowest free-energy state. The distribution of ``touch-points'' 
defines a spatial map where the energy of the {\bf K} points vary. In effect,
unlike the perfect 2D sheet, the graphene on the substrate has a $k_z$
bandstructure which is a set of strongly localized (flat) states, with an average
extension in the z-direction of $\sim z_g$ at the touch points.
 Thus the square of the wavefunction of an electron in graphene, projected
 on the $z$ - direction is  {\it no longer} a $\delta$-function
 $\delta(z-z_g)$.

\begin{figure}
\includegraphics*[width=8.0cm,height=10.5cm]{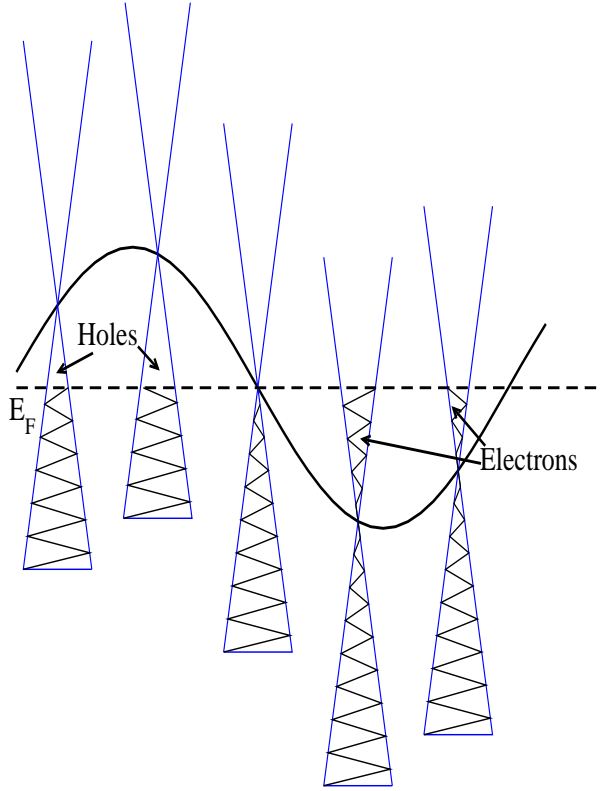}
\caption
{(Color online)Position dependent Dirac-Weyl energy-dispersion cones arising
from a single Fourier componet of a wavy graphene sheet (shown as a thick black curve),
 and the
resultant pockets of electrons and  holes.  These are defined by the common
 electrochemical potential 
of the system}
\label{wavycones}
\end{figure}

The above picture translates to the real space with the 
energy $\epsilon_K$ at the {\bf K} points acquiring a 
a small $z$-dependence, while the global Fermi energy of the system remains
constant and aligned to the chemical potential. That is, for any $k$
measured with respect to {\bf K}, we have
\begin{eqnarray}
\epsilon(k,z)&=&\varepsilon_F(z)+\epsilon(k)\\
\epsilon(k)&=&\pm v_F|k|
\end{eqnarray}
In Fig.~\ref{wavycones} we show such an array of $z$-dependent energy cones
for graphene, for a single Fourier component of the structure factor of
the graphene sheet. A sum over all such Fourier components is 
equivalent to a weighted sum over all  possible folded bands
in the reciprocal-space picture. We show only a single cone for the
two degenerate valleys having a spin degeneracy as well. That is,
we assume that the {\bf K}, {\bf K$^\prime$}
degeneracy is negligible compared to the 
variation of the absolute value of $\epsilon(K)$ at the 
length scale $L_{xy}$, as the former
corresponds to length scales an order of magnitude smaller.
Although the {\bf K}-point energy of the cones bob up and down,
 the system will have a
common electrochemical potential which defines the energy transport
across the system when a potential is applied. Clearly, when $V_g$
drops below $\varepsilon(l_z)$ the system becomes insensitive to
$V_g$, and hence we can identify $v_m=\varepsilon(l_z)$. Below this
threshold the carrier density is determined by the waviness of the
graphene sheet and not by the applied voltage which is smaller than
$v_m$.
 Given similar methods of preparation and similar conditions of
observation, the $l_z/L_{xy}$ parameter and the resultant carrier
density would be approximately ``universal'' in the sense that it
would lie within a limited range of possible values.

If the average $z$-extension of the electron is $l_z$, its
possible energy states would be quantized with a spectrum of
``particle in a box''  like energies $\varepsilon_n(l_z)$. The lowest of these
would correspond to the $n=1$ state of some effective
 minimum value of the z-direction
 parameter. A number of such states would be occupied, depending on $V_g$.
Once $V_g$ falls below $v_m$, one may assume that only these lowest
states, i.e., $n=1$, in each cone
 at $-v_m+v_Fk$ or $+v_m-v_Fk$ would be occupied
by the respective carriers. Thus we see that this system can indeed
reduce to just a minimal conduction channel where each channel
has a spin and valley degeneracy unless external magnetic or
electrostatic fields are present. This argument is independent
of the exact extent of $l_z$ and $L_{a}$.
The z-direction bandstructure defines channels to which the Landauer-Buttiker
formula \cite{lanbut} can be applied.
Taking the  spin and pseudospin
degeneracy into account,
 a minimum conductance of
$4(e^2/h)$ would be expected for these systems. That is, although the
value of the threshold $v_m$ would {\it not} be universal, the conductance
will have a minimal universal value corresponding to the occupation
of the lowest $z$-quantized state.
Note that this picture is quite different to the mobility-edge conductance model
of disordered semiconductors\cite{fradkin}, and closer to a model of a quasi-2D
electron gas \cite{quasi2d} where the thickness varies from point to
point.\\

\section{Conclusion}
In this study we have shown that simple models of the silicon dioxide
substrate can be constructed and used to give some insight into the
substrate/graphene interaction. These models show that if the substrate is
free of dangling bonds, a graphene layer positioned at about 4-5 \AA$\,$ away from
the substrate preserves its DW bands. However, the energy at the {\bf K}
point acquires a spatial variation. The atomically thin 2-D layer effectively
acquires an $z$-extension, and the $z$-confined states in this quasi-2D
system provide the quantization necessary to define a minimum conductance
of $\sim 4e^2/h$. This behaviour is {\it different} from
what is expected of ideal graphene where
the conductance should go to zero when the gate voltage goes to zero.

\end{document}